\documentclass[aps,amsmath]{revtex4}
\usepackage{graphicx}
\def\Journal#1#2#3#4{{#1} {\bf #2}, #3 (#4)}

\def\NPA{{Nucl. Phys.} A}

\def\PRL{Phys. Rev. Lett.}

\def\be{\begin{equation}}
\def\ee{\end{equation}}
\def\bea{\begin{eqnarray}}
\def\eea{\end{eqnarray}}

\def\beq{\begin{equation}}
\def\eeq{\end{equation}}
\def\bea{\begin{eqnarray}}  \def\eea{\end{eqnarray}}
\def\lsim{\raise0.3ex\hbox{$<$\kern-0.75em\raise-1.1ex\hbox{$\sim$}}}
\def\gsim{\raise0.3ex\hbox{$>$\kern-0.75em\raise-1.1ex\hbox{$\sim$}}}
\def\1{{\rm 1\mskip-4.5mu l} }
\newcommand{\noi}{\noindent}
\begin{document}
\def\thepage{\roman{page}}
\title{Strangeness Production in STAR at RHIC}

\author{Christelle Roy for the STAR Collaboration}

\affiliation{SUBATECH \\ UMR 6457 IN2P3/CNRS, Ecole des Mines de Nantes, Universit\'e de Nantes \\
La Chantrerie BP20722 44307 Nantes cedex 3, France} 

\begin{abstract}
Strangeness study is one of the major goal of the STAR 
experiment at RHIC. Results presented here have been obtained from
analyses restricted to the mid-rapidity region of Au-Au collisions 
at $\sqrt{s_{NN}}$ = 130 GeV. An onset of the understanding of the quark
matter behavior produced at this relativistic energy can be
stressed from the investigation of the relative and absolute yields of
strange particle production, their transverse mass distributions and
from comparisons with previous results obtained by AGS and SPS experiments
\end{abstract}

\maketitle

\section{Strange proofs for a Quark Matter Investigation}
\noi It is well known now that studying strangeness should provide
information of different nature about the quark matter which has been
created during the heavy ion collision. Since severals years, it is
proposed that the quark gluon plasma (QGP) formation can be revealed
via the observation of an enhancement in the strangeness production
compared to a "normal" yield in a hadronic gas~\cite{ra}. However,
from recent results presenting by the NA49 collaboration~\cite{va}
measuring strangeness production in proton-proton (pp), proton-nucleus
(pA) and nucleus-nucleus (AA) reactions, one learns that the
definition of this enhancement has to be defined very rigorously and
precisely, since an enhancement is already seen in pA with respect to
pp collisions. Systematical studies have to be drawn up as a function
of the collision energy, the system size, the degree of centrality of
the reaction and even the particle species.

\noi Upstream from the QGP highlight, an issue is to understand how
strangeness is produced. Relativistic heavy ion physics could be
understood with the characterization of some fundamental points like : \\
$\_$ original environment (baryon density, degree of stopping) by
measuring antibaryon/baryon ratios. \\
$\_$ production mechanisms which can be assessed from different approaches :
the comparison of spectra (in terms of transverse mass or rapidity) of
particles and their anti-particles can inform if species obey or not 
the same production mechanisms depending on 
similarities or discrepancies which can be seen between them; 
ratios can also be informative on   
mechanisms such as coalescence processes for example. \\
$\_$ relative and absolute yields of strange particle production,
especially those of kaons since they carry the majority of the
strangeness quark content of the reaction. \\
$\_$ time scale of the different phases encountered during the system
evolution by the observation of the production or suppression of
resonances.\\
$\_$ amplitude of the rescattering or collective effects by looking at
the transverse momentum spectra.\\
$\_$ degree of suddenness of the hadronisation, depending on the
coincidence or not of the chemical and thermal freeze-out parameters
extracted from particles ratios.

STAR (Solenoidal Tracker At Rhic) constitutes a very well suited
experiment for strangeness measurement. It includes indeed a large
tracker apparatus, the Time Projection Chamber (TPC) with a large 
p$_T$-y acceptance and a full azimuthal coverage. 

\section{Some experimental considerations}
\noi Results presented in the following have been obtained from analyses of
events collected during the year 2000. RHIC delivered gold
beams at an energy of $\sqrt{s_{NN}}=$130 GeV. At this time, the TPC
was the main detector of STAR experiment~\cite{st} operating with a
0.25 T field. The TPC allows for tracking of charged particles emitted in
the $\mid\eta\mid$=1.8 and $p_T\geq$0.75 GeV/c coverage for a
collision occurring at the center of the detector. The event centrality
is selected according to the multiplicity distribution of the
negatively charged particles~\cite{ac}, once the event reconstruction
is done. The number of events which can be used for physics analyses is
equal to 460k and 330k for minimum bias and the most
central events (top 5$\%$) respectively. Minimum bias events are triggered by
requiring a coincidence of signals in both Zero Degree Calorimeters which 
detect spectator neutrons. A high threshold in the Central Trigger Barrel 
is set for selecting the central events.

\noi More specifically, several methods are used for the identification of
strange particles. 
Strange particles (K$^0_s$, $\Lambda$, $\Xi$, $\Omega$) 
decaying at a certain distance from the primary vertex of the collision, are 
reconstructed using a topological method. The tracks of their
daughters are reconstructed in the TPC, extrapolated back towards a
common origin and the kinematics of the parent are calculated.

\noi Charged kaons can be identified from their energy loss in the
TPC if their momentum is lower than 0.6 GeV/c. Beyond this limit,
their energy losses can not be separated unambiguously from those of
the pions due to the TPC resolution. A topological
method, the so-called kink method, is used for the determination of
charged kaons via the K$\rightarrow\mu\nu$ channel. In this case, the
charged kaon and charged daughter are used to reconstruct the
kinematics of the parent. Opposite to the dE/dx measurement,
the kink method allows to avoid any restriction on the momentum of the
kaons and thus to identify them up to 2 GeV/c.

\noi Reconstruction of resonances ($\phi$, K$^{0*}$) is achieved 
by combinatorics calculating the invariant mass with all permutations
of candidate decay particles. The background is obtained by the
mixed-event technique. The combinatorics can be applied also for the
reconstruction of the strange baryons, thus allowing an effective 
cross-check with the topological method.

\vspace*{0.2cm}
\noi Note that results (excepted anti-baryon over baryon ratios) presented
here are corrected for tracking efficiencies, acceptance and detector
effects. Besides, as a first attempt, the analyses have been
restricted to the mid-rapidity region.

\section{What rules strangeness production ?}
\subsection{Production yields}
\noi Kaons carry about 70$\%$ of the strange quark content created during
the collision, hence providing a good estimate of the amount of
produced strangeness. The yields of kaons per unit of rapidity in the
most central collisions reach for the K$^+$, K$^-$ and K$^0_s$
35$\pm$3(stat)$\pm$5(syst), 30$\pm$3(stat)$\pm$4(syst) and
35.1$\pm$0.6(stat) respectively~\cite{ca}. Results are similar one to
each other and when compared to the multiplicity of the
negatively primary charged particles~\cite{ad} (280$\pm$1(stat)$\pm$20(syst)),
they demonstrate the large amount of strangeness created at RHIC (more
than 10 times the amount found in Pb-Pb collisions at SPS).

\noi Analyses of the $\Phi$ mesons reveal also a large production of
strangeness~\cite{phi}. The energy dependence of the $\Phi$/h$^-$ ratio in heavy
ion collisions from $\sqrt{s_{NN}}$ = 5 up to 130 GeV indicates that this
ratio increases with the collision energy hence that the $\Phi$
production increases faster than that of the h$^-$, up to RHIC energy.

\begin{figure}
\begin{center}
\begin{minipage}{0.45\linewidth}
\includegraphics[height=\linewidth,width=\linewidth]{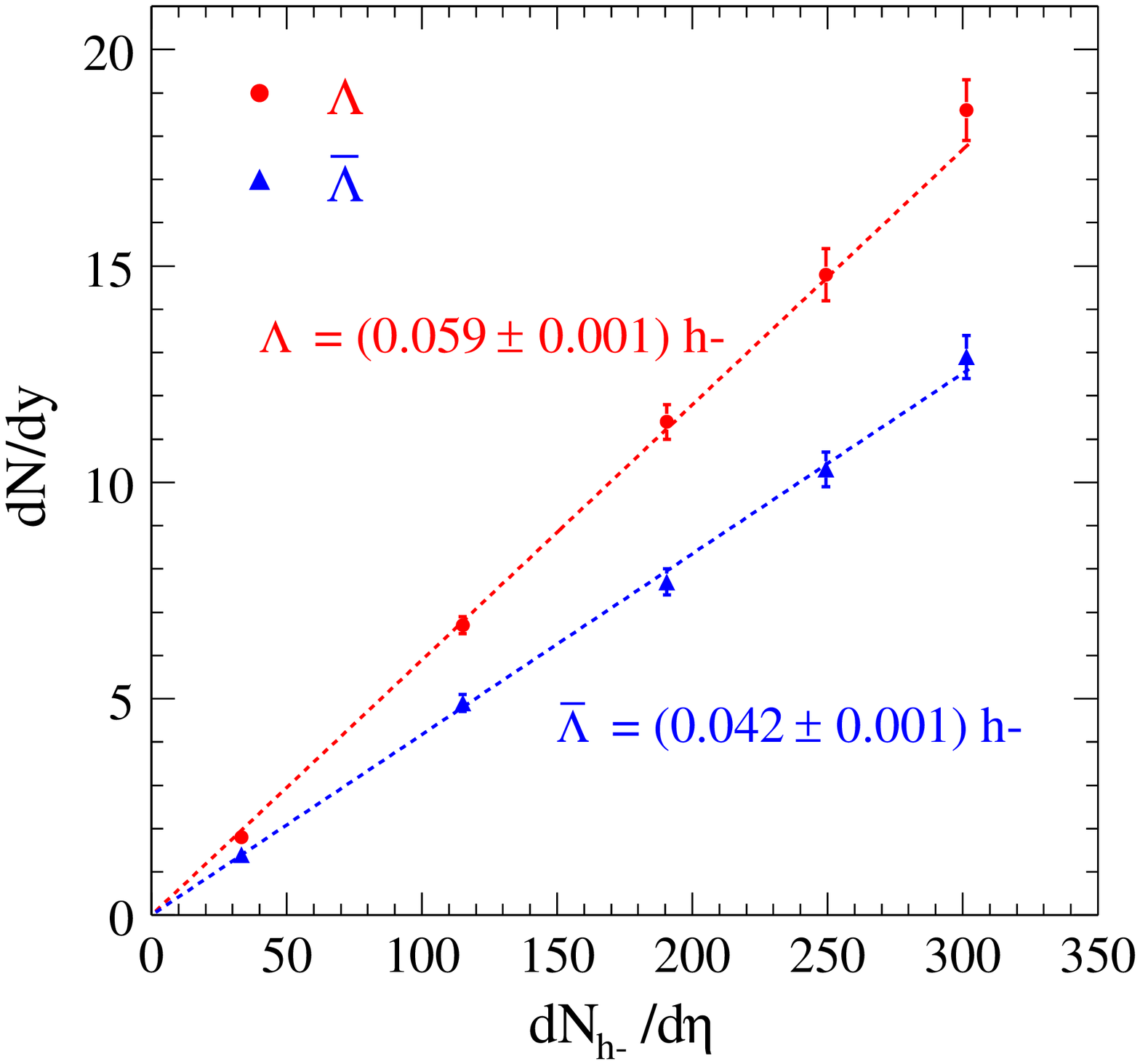}
\caption{$\overline\Lambda$ and $\Lambda$ yields versus the 
negative hadron (h$^-$) multiplicity at mid-rapidity. Errors are 
statistical only and hyperon yields not feed-down corrected}
\label{fig:lambda1}
\end{minipage}
\hspace{\fill}
\begin{minipage}{0.45\linewidth}
\includegraphics[height=\linewidth,width=\linewidth]{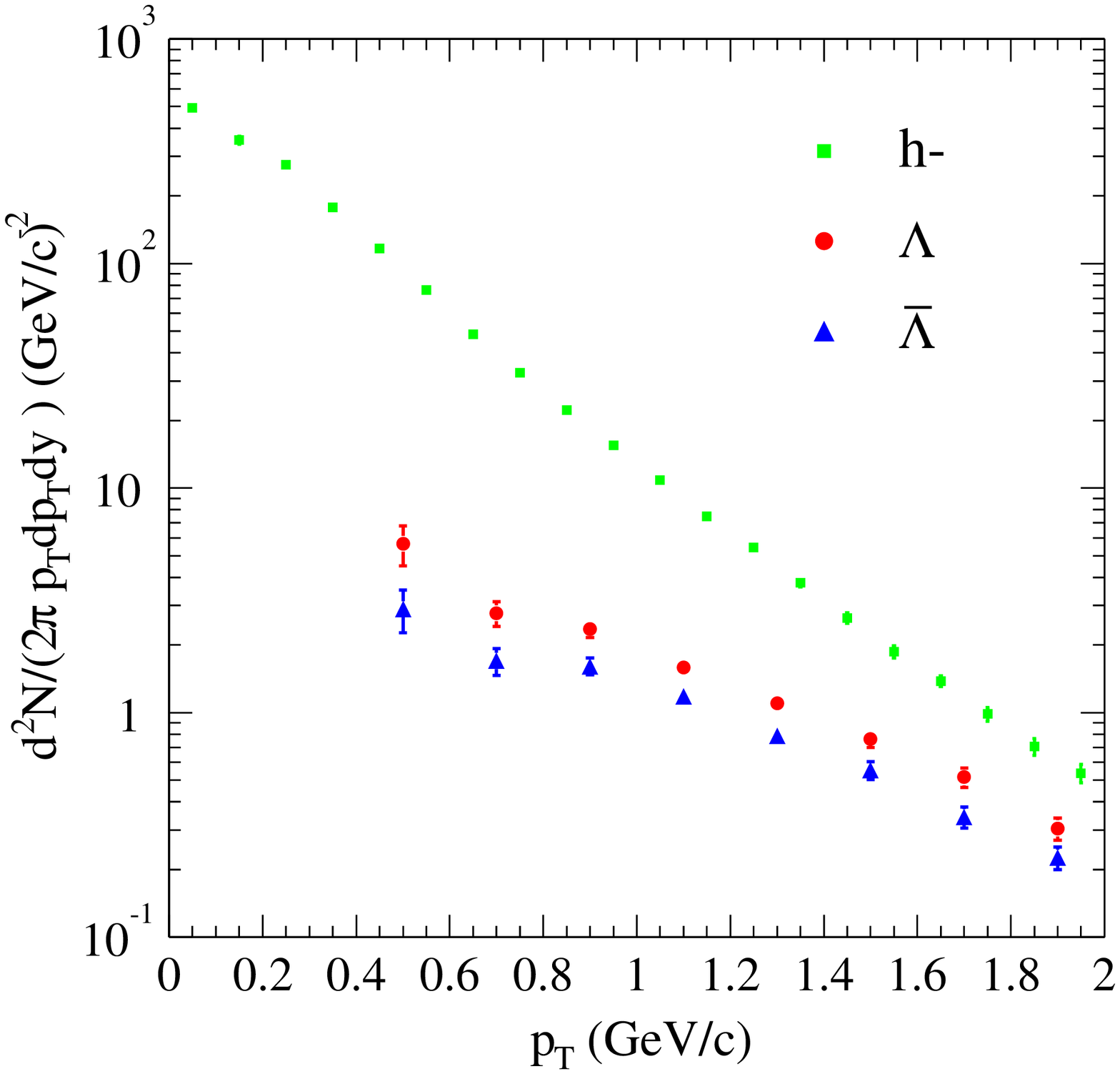}
\caption{$\overline\Lambda$, $\Lambda$ and negative hadron (h$^-$) 
transverse momentum distributions for the $5\%$ most central collisions}
\label{fig:lambda2}
\end{minipage}
\end{center}
\end{figure}

The $\Lambda$ and $\overline\Lambda$ production at mid-rapidity is
shown on figure~\ref{fig:lambda1} as a function of the negative
hadron (h$^-$) multiplicity. The hyperon production increases with the
centrality of the collision reaching the values of 18.6$\pm$0.7(stat.)
and 12.9$\pm$0.5 for respectively $\Lambda$ and $\overline\Lambda$ in
the 5$\%$ most central collisions. Moreover, this hyperon production
appears to be linearly proportional to that of the h$^-$. 
Figure~\ref{fig:lambda2} exhibits clearly that the p$_T$ distributions
of the $\overline\Lambda$ and $\Lambda$ are much flatter than the
p$_T$ distribution of the negatively charged particles. The hyperon
production becomes very important at high p$_T$ suggesting that the
baryon to meson ratio should exceed 1 at very high p$_T$ values. This
surprising trend has been observed also by the PHENIX collaboration
measuring $\overline{p}$/$\pi^-$ ratio~\cite{ja}. The interpretation
of such a behavior is still an open question : Shuryak mentioned
simple flow effects~\cite{shu} while Vitev and Gyulassy propose a
novel non-perturbative component of baryon dynamics~\cite{gy} as an
explanation.

\noi The $\Xi$ and $\overline\Xi$ yields correspond, for the 14$\%$ most
central collisions in the mid-rapidity region, to 3.07$\pm$0.13(stat) and
2.63$\pm$0.12(stat), respectively. 

\noi Due to an insufficient number of events, the absolute yields of
$\Omega$ and $\overline\Omega$ are not yet accessible despite the
signal of the $\Omega + \overline{\Omega}$ system has been extracted
with a very good signal over noise ratio (S/N = 3.6). The new data of
the year 2001 will allow solving this statistics issue.

\subsection{Initial environment}
\begin{figure}
\begin{center}
\begin{minipage}{0.45\linewidth}
\includegraphics[height=\linewidth,width=\linewidth]{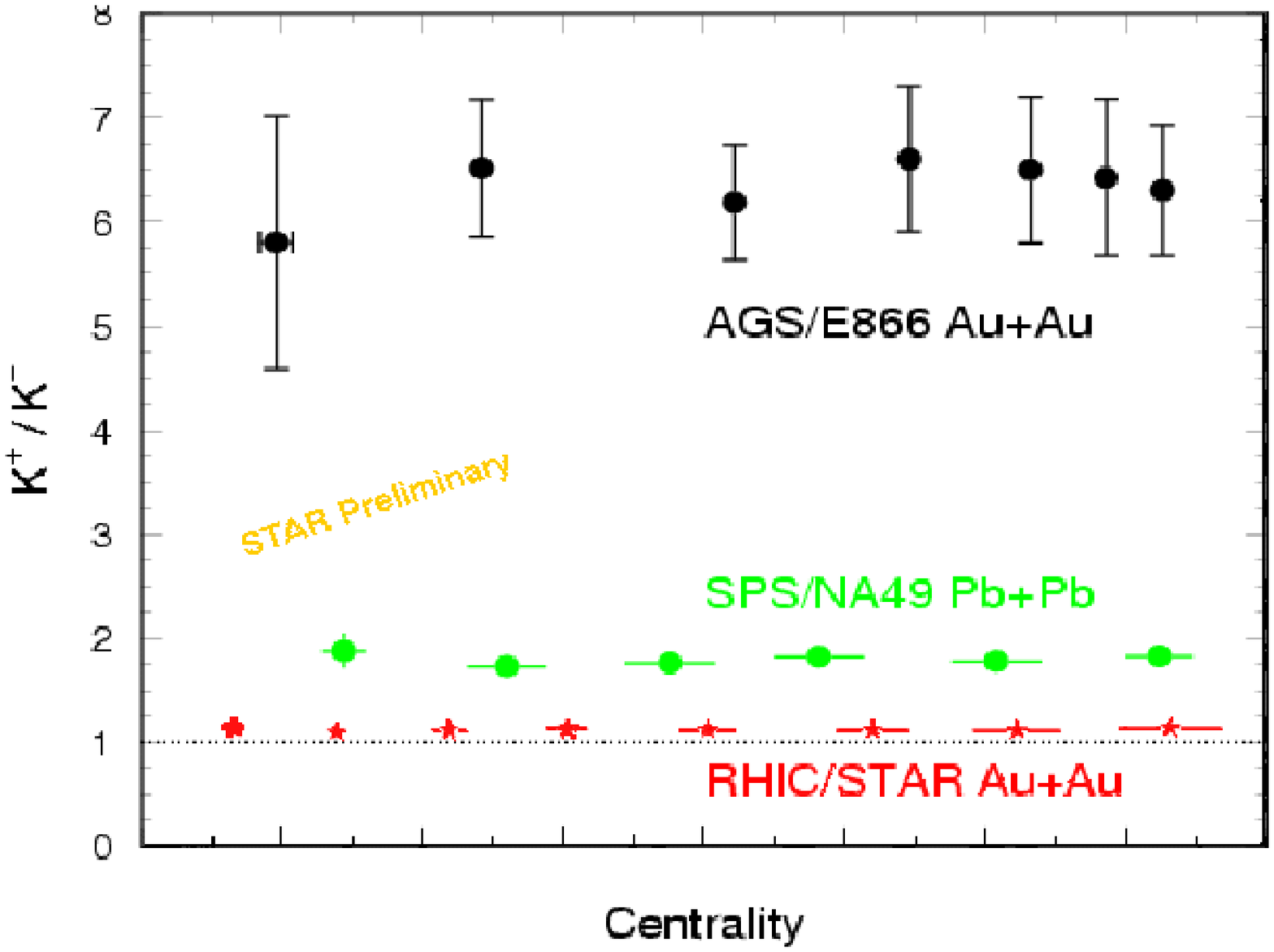}
\caption{K$^+$/K$^-$ ratio as a function of the centrality, at AGS(E866), 
SPS(NA49) and RHIC(STAR) energies}
\label{fig:kaons}
\end{minipage}
\hspace{\fill}
\begin{minipage}{0.45\linewidth}
\includegraphics[height=\linewidth,width=\linewidth]{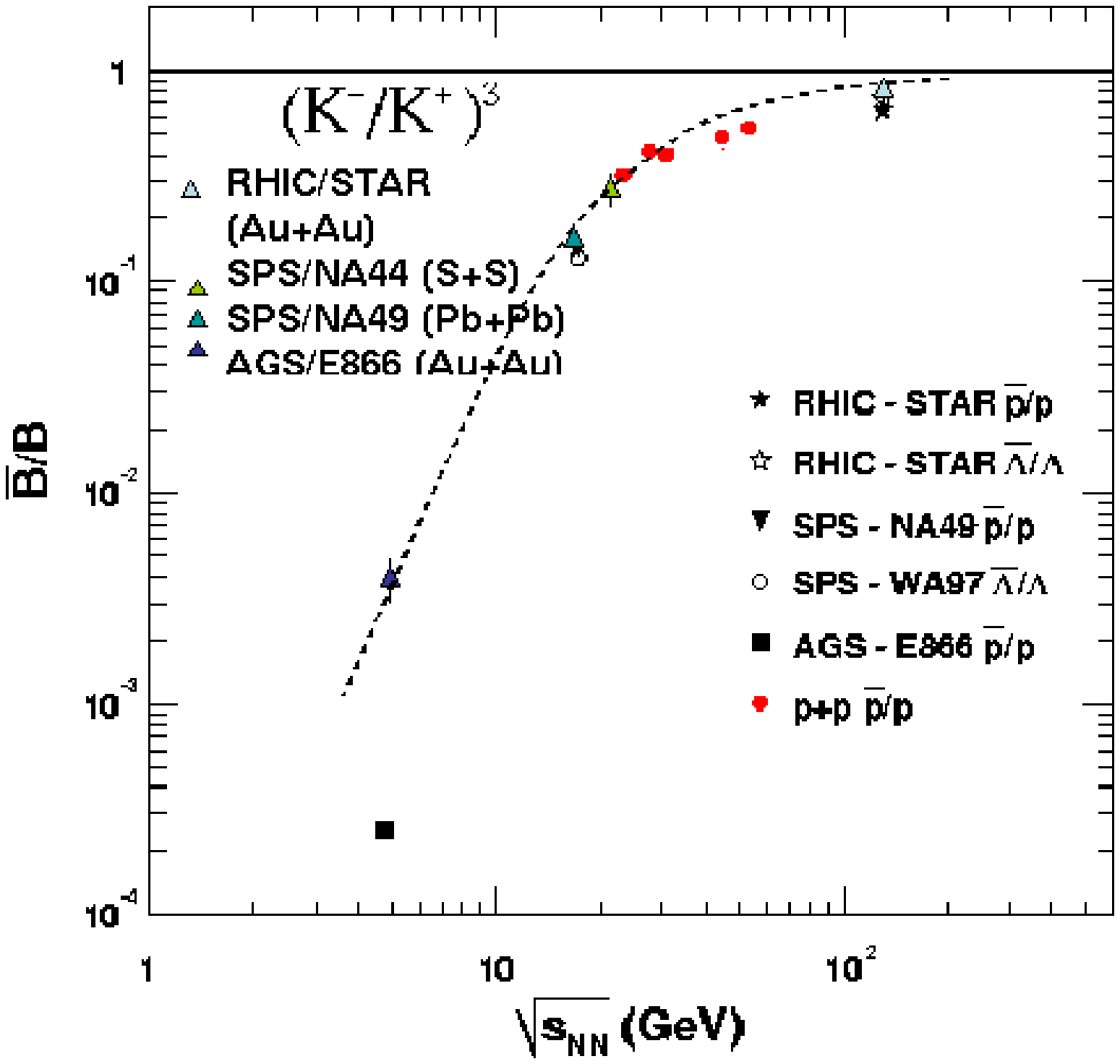}
\caption{Anti-baryon to baryon ratios excitation function. Triangles correspond 
to (K$^-$/K$^+$)$^3$ ratios - dashed line is just a guide for the eye.}
\label{fig:exfunc}
\end{minipage}
\end{center}
\end{figure}
Due to the respective quark content of K$^+$ and K$^-$, their ratio
indicate the amplitude of the net-baryon
density. Figure~\ref{fig:kaons} presents the K$^+$/K$^-$ ratio at
mid-rapidity as a function of the centrality. It is 
equal to 1.071$\pm$0.008(stat) and is rather constant as a function of
the centrality as observed at AGS and SPS energies. The decrease of
this ratio as a function of the collision energy reflects the change
and drop of the net-baryon density as the energy increases.

\noi The excitation function of anti-baryon to baryon ratios
(figure~\ref{fig:exfunc}) renders the strong increase anti-baryon
production going from the lower to higher energy. Even at RHIC the
net-baryon density is still not equal to 1 or, differently speaking,
system is still not baryon free despite the important drop of the
degree of stopping going from AGS to SPS to RHIC.

\begin{figure}
\begin{center}
\begin{minipage}{0.45\linewidth}
\includegraphics[height=\linewidth,width=\linewidth]{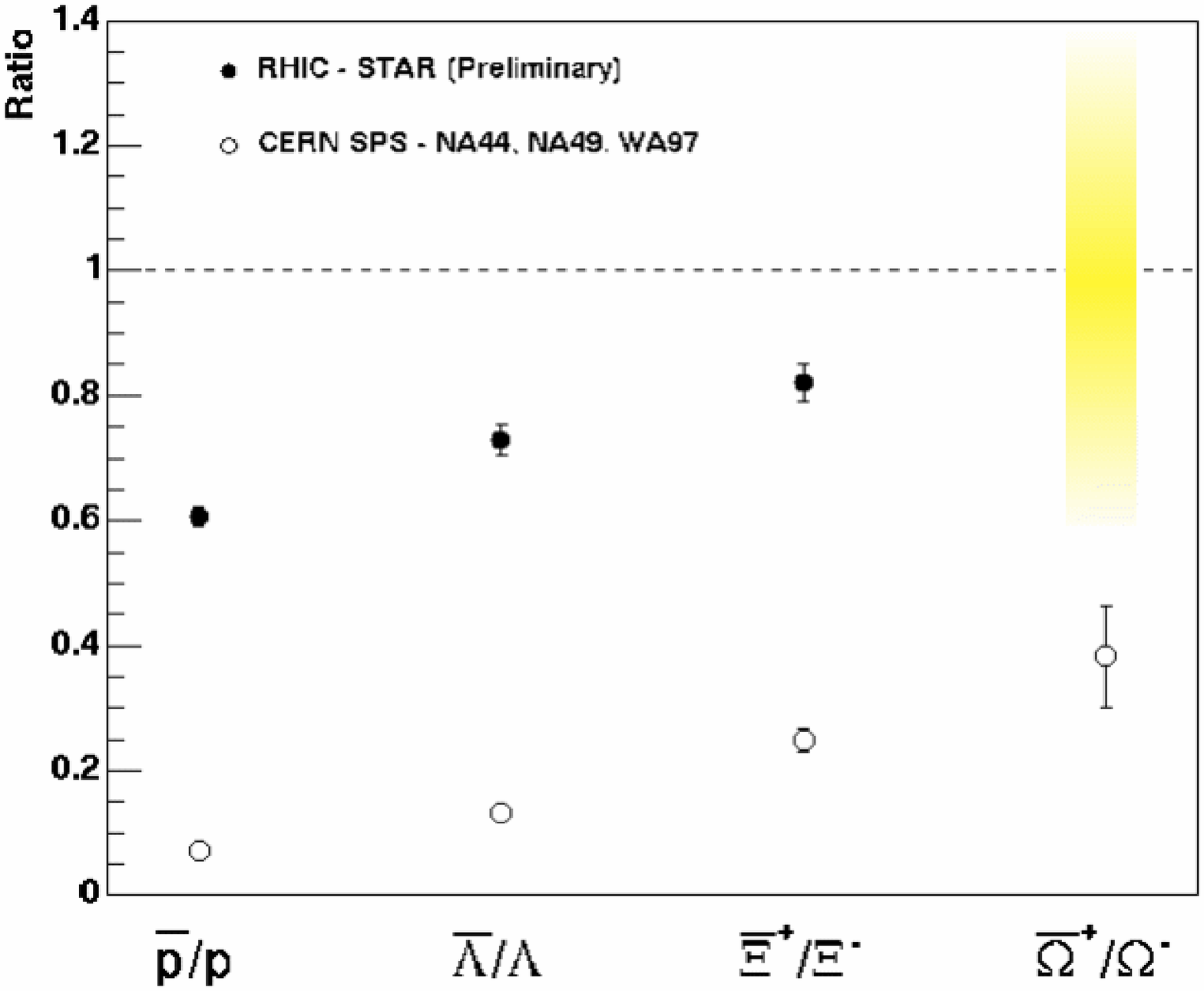}
\caption{$\overline{B}$/$B$ ratios as a function of the strangeness 
content at SPS and RHIC energies.}
\label{fig:stran}
\end{minipage}
\hspace{\fill}
\begin{minipage}{0.45\linewidth}
\includegraphics[height=\linewidth,width=\linewidth]{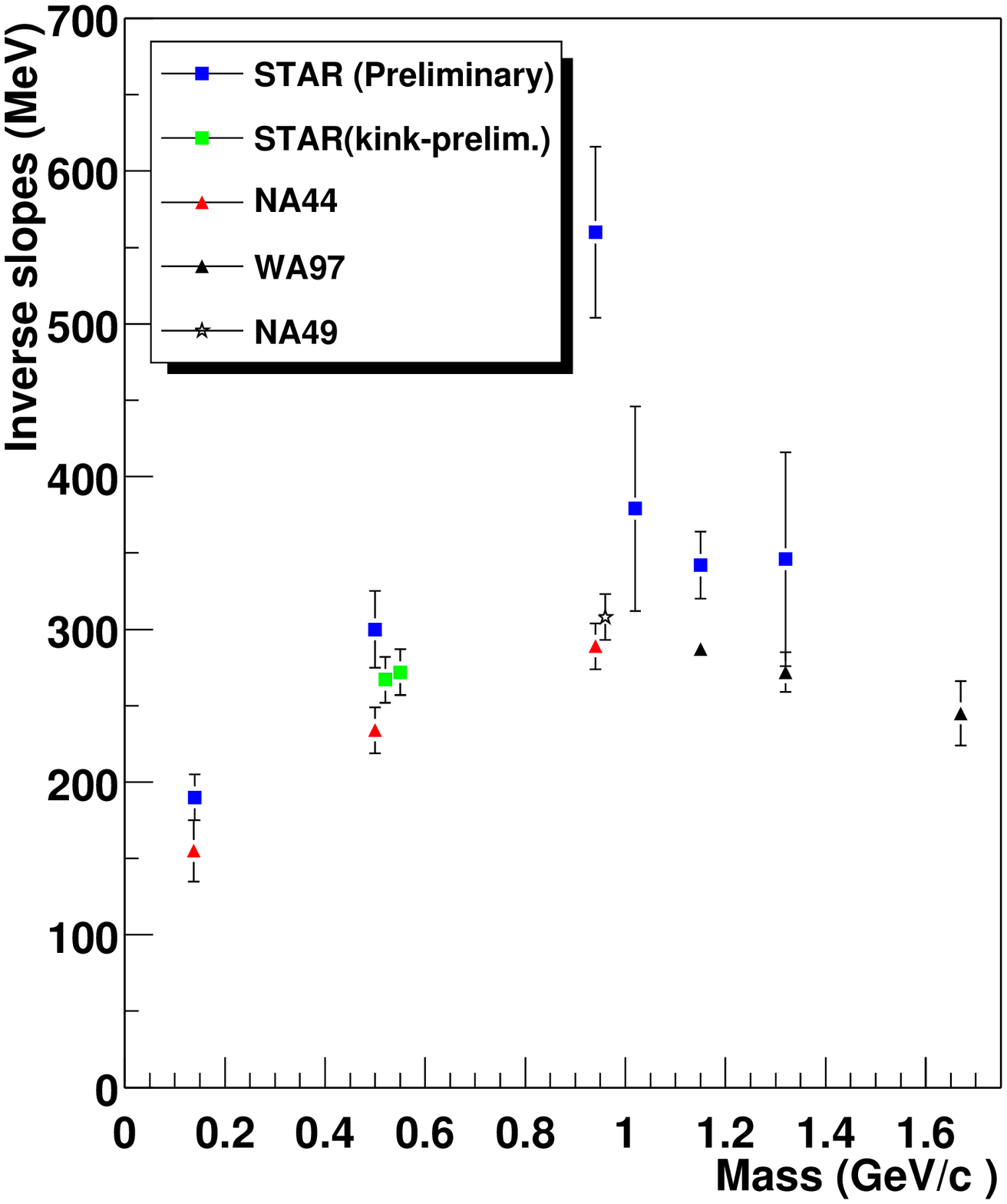}
\caption{Inverse slopes of transverse mass spectra versus  
particle mass, at SPS and RHIC energies.}
\label{fig:flow}
\end{minipage}
\end{center}
\end{figure}

\noi Figure~\ref{fig:stran} presents ratios of antibaryons to baryons 
according to their strangeness content for SPS and RHIC data. As
predicted~\cite{ra} and also already seen at SPS, STAR measurements
indicate the enhancement of the ratios as strangeness content
increases:

\hspace*{2.cm}$\overline{p}/p = 0.63\pm0.02\mbox{$\;$(stat)}\pm0.06\mbox{$\;$(syst)}$ 
\hspace*{0.5cm}$\overline{\Lambda}/\Lambda = 0.73\pm0.03\mbox{$\;$(stat)}$

\hspace*{4.cm}$\overline{\Xi}/\Xi = 0.83\pm0.03\mbox{$\;$(stat)}\pm0.05\mbox{$\;$(syst)}$

\noi It is difficult to rule on the $\overline{\Omega}/\Omega$ ratio :
despite it seems to be compatible with 1, there is still large
statistical errors to allow an accurate evaluation.  The data
collected during 2001 should allow a more precise measurement.

\noi  From these ratios, Rafelski {\em et al.}~\cite{fo} have extracted the
 values of chemical potential and temperature at which hadronisation
 occurs. From $\overline{p}$/p value~\cite{pp}, they estimate that the
 baryo-chemical potential is equal to 32(38)MeV if the temperature is
 fixed at 150(175)MeV, respectively. Furthermore, from
 $\overline{\Lambda}$/$\Lambda$ and $\overline{\Xi}$/$\Xi$ ratios,
 they show that the strange quark fugacity is consistent with 1 and
 according to their thermal model, it corresponds to the value
 expected for sudden hadronisation.

\noi On the other hand, Braun-Munzinger and collaborators have performed
a fit to the preliminary RHIC data~\cite{pbm} leading to a 
temperature of 175$\pm$7MeV and a baryo-chemical potential of
51$\pm$6MeV. 

\noi According to statistical models, the fireball seems to have reached a
high degree of chemical equilibration. Note however, that to be
conclusive, such an approach should be investigated with data
extrapolated to a 4$\pi$ coverage.

\subsection{Production mechanisms}
  
\noi The $\overline p$/$p$ ratio indicates that, contrarily to
AGS~\cite{ags} or SPS~\cite{sps1} trends, pair processes dominate baryon
transport at RHIC, by about a factor 2 if one assumes that
anti-proton production is due to pair processes (production and
annhililation included) while protons come from both baryon transport
and pair processes. Thus, 2/3 of protons come from pair processes.

\noi Furthermore, ratios seem to be consistent with simple quark
coalescence model~\cite{coa}. This latter assumes that  quark
matter hadronizes via a sudden recombination of its quark and
anti-quark constituents . Within this approach, $\overline{B}$/B
ratios can be predicted from a simple quark counting :

$\qquad\qquad\qquad \frac{\overline{\Lambda}}{\Lambda} = D * \frac{\overline{p}}{p} 
\quad \mbox{and} \quad \frac{\overline{\Xi}}{\Xi} = 
D * \frac{\overline{\Lambda}}{\Lambda} = D^2 *\frac{\overline{p}}{p} 
\quad\mbox{...}\quad\mbox{where}\quad D = \frac{K^+}{K^-}$

\noi Thus, the D values calculated with the various baryon ratios measured
by STAR, are very close to the D value directly measured
(1.071$\pm$0.008(stat)). The similarities between calculated and
measured D values indicate the consistency of the coalescence model
predictions with the baryon ratios at RHIC, as already seen at
SPS energies.

\noi Considering again the K$^-$/K$^+$ ratio (rather than K$^+$/K$^-$
ratio), the excitation function of the $\overline{B}$/B can be
revisited. Indeed, at energies far above the kaon production
threshold, one can assume that the relative abundance of strange and
anti-strange quarks do not impact the relative production of K$^-$ and
K$^+$. K$^-$/K$^+$ ratio becomes similar to $\overline{u}$/u. On the
other hand, one can postulate that $\overline{u}/u \sim
\overline{d}/d$, which is reasonable once again at high energies for
which the isospin difference at mid-rapidity is less significant. This
implies the following relation : $\overline{B}/B
\sim (\overline{u}/u)^3 \sim (K^-/K^+)^3$. The curve on 
figure~\ref{fig:exfunc} linking up the $(K^-/K^+)^3$ data at various
$\sqrt{s_{NN}}$ indicates that the same behavior is observed for
K$^-$/K$^+$ and $\overline{p}$/p at the highest energies. The
different behavior seen at AGS can be explained by the fact that, at this
lower energy, the assumptions are less valid and also the
absorption of $\overline{p}$ in the medium is large ($\sim$20$\%$).

\subsection{Dynamics}

\noi Transverse mass spectra have been investigated for strange hadrons.
For each couple (K$^+$-K$^-$, $\Lambda$-$\overline{\Lambda}$ and
$\Xi$-$\overline{\Xi}$), the m$_T$ distributions are rather identical for
particles and anti-particles (same slopes), informing that they
present the same final state spectra, despite the
processes producing (anti-)particles may be very different.

\noi The values of inverse slope extracted from a fit to the experimental transverse
mass distributions with an exponential function are summarized on
figure~\ref{fig:flow} as a function of the particle masses. It is
well-known that within a hydrodynamical approach, these slopes can be
interpreted as to be due to the combination of a thermal component
(the corresponding parameter being the temperature at freeze-out) and
a collective one (the parameter being the flow velocity). The
comparison of STAR and SPS~\cite{sps2} data allows to highlight two
facts. First, it appears clearly that there is a strong
indication of radial flow at STAR. Second, strange particles follow
the same behavior at SPS and RHIC energies namely that the inverse
slopes decrease with increasing mass.  A hydrodynamical
approach~\cite{hydro} leads to the following parameters : T$_{FO}$ =
130 MeV and $\beta_{flow}$ = 0.5 c.  Nevertheless, it is not trivial
to obtain presently a coherent picture from the RHIC measurements (see
$\overline{p}$ or $\phi$ points). A possible explanation is that the
slopes depend strongly on the transverse momentum range where the fit
is performed (more details can be found in ~\cite{jo}) : For example,
in the case of the $\Lambda$, a unique exponential function is not
able to reproduce the distribution. Investigations have to be pursued.

\section{Attempt of answers}
\noi STAR has highlighted the important amount of strangeness produced at
RHIC energy. Information has been obtained about the original environment of the
particle production. Measurements of particle ratios  lead to the
conclusion that the net-baryon density drops with the decrease of the
degree of stopping going from AGS, to SPS to RHIC but even at the
higher energy, the system is still not baryon free.

\noi Ratios appear to be consistent with a simple coalescence model. They
weakly depend on the transverse momentum, suggesting that the
rescattering is not significant. It has also been observed that baryon
and anti-baryon present similar final state spectra however  
mechanisms of their production can be different and further analyses
have to be done. Inverse slopes of transverse mass distributions exhibit a strong
radial flow governing the fireball evolution at RHIC. And it becomes
quite puzzling to observe that statistical approach is able to
reproduce quite well the data. 

\noi STAR future analyses will be very promising for the strangeness
investigation. RHIC is running at full energy ($\sqrt{s_{NN}}$ = 200
GeV), statistics will increase and an important issue is the
adding of the Silicon Vertex Detector to the TPC. 


\end{document}